\documentclass[10pt,journal]{IEEEtran}

\ifCLASSINFOpdf
	\usepackage[pdftex]{graphicx}
\else
	\usepackage[dvips]{graphicx}
\fi
 \usepackage{ifthen}
 \newboolean{bookver}
 \setboolean{bookver}{false}

\usepackage[latin1,utf8]{inputenc} 
\usepackage[sort&compress]{natbib}
\usepackage[cmex10]{amsmath} 
\usepackage{array} 
\usepackage{amssymb} 
\usepackage{amsthm} 
\usepackage{dsfont} 
\usepackage{bm} 
\usepackage{acro}
\usepackage{color} 
\usepackage[pdftex]{hyperref}
\usepackage[all]{hypcap}
\usepackage{cleveref}
\usepackage{epstopdf}
\usepackage{multirow}
\usepackage{booktabs}
\usepackage{bigstrut}

\usepackage{enumitem}
\usepackage{tabu}
\usepackage{multirow}
\usepackage{subfig}

 \bibpunct{[}{]}{,}{n}{,}{;}

\definecolor{myBlue}{rgb}{0,0,0.55}
 \definecolor{xBlue}{rgb}{0,0,0.5}
 \hypersetup{
  linkcolor=  myBlue,
  citecolor=  myBlue,
  urlcolor=   myBlue,
  pdffitwindow=       true,
  breaklinks=         true,
  pdfcenterwindow=    true,
  pdfstartview=       FitH,
  bookmarks=          false,
  bookmarksopen=      false,   
  bookmarksnumbered=  false,   
  pdfhighlight=       /P,
  linktocpage=        true,   
  colorlinks=         false,
  pdfborder=          0 0 0,
  pdftitle=           {Network Architectures for the Society in Motion},
  pdfauthor=          {Stefan Schwarz and Markus Rupp},
  pdfkeywords=        {},
  }

\usepackage[switch]{lineno}
\newcommand*\patchAmsMathEnvironmentForLineno[1]{%
 \expandafter\let\csname old#1\expandafter\endcsname\csname #1\endcsname
 \expandafter\let\csname oldend#1\expandafter\endcsname\csname end#1\endcsname
 \renewenvironment{#1}%
    {\linenomath\csname old#1\endcsname}%
    {\csname oldend#1\endcsname\endlinenomath}}%
\newcommand*\patchBothAmsMathEnvironmentsForLineno[1]{%
 \patchAmsMathEnvironmentForLineno{#1}%
 \patchAmsMathEnvironmentForLineno{#1*}}%
\AtBeginDocument{%
\patchBothAmsMathEnvironmentsForLineno{equation}%
\patchBothAmsMathEnvironmentsForLineno{align}%
\patchBothAmsMathEnvironmentsForLineno{flalign}%
\patchBothAmsMathEnvironmentsForLineno{alignat}%
\patchBothAmsMathEnvironmentsForLineno{gather}%
\patchBothAmsMathEnvironmentsForLineno{multline}%
}

\definecolor{revgreen}{rgb}{0,.75,0} 

\DeclareAcronym{3D}{
  short = 3D,
  long  = three dimensional
}
\DeclareAcronym{3GPP}{
  short = 3GPP,
  long  = Third Generation Partnership Project
}
\DeclareAcronym{4G}{
  short = 4G,
  long  = fourth generation
}
\DeclareAcronym{5G}{
  short = 5G,
  long  = fifth generation
}
\DeclareAcronym{AAS}{
  short = AAS,
  long  = active antenna system
}
\DeclareAcronym{BLER}{
  short = BLER,
  long  = block-error ratio
}
\DeclareAcronym{BSM}{
  short = BSM,
  long  = basic safety message
}
\DeclareAcronym{CAM}{
  short = CAM,
  long  = cooperative awareness message
}
\DeclareAcronym{C-ITS}{
  short = C-ITS,
  long  = cooperative ITS
}
\DeclareAcronym{CoMP}{
  short = CoMP,
  long  = coordinated multipoint transmission
}
\DeclareAcronym{CP}{
  short = CP,
  long  = cyclic prefix
}
\DeclareAcronym{C-RAN}{
  short = C-RAN,
  long  = cloud radio access network
}
\DeclareAcronym{CSI}{
  short = CSI,
  long  = channel state information
}
\DeclareAcronym{CSIR}{
  short = CSIR,
  long  = channel state information at the receiver
}
\DeclareAcronym{CSIT}{
  short = CSIT,
  long  = channel state information at the transmitter
}
\DeclareAcronym{D2D}{
  short = D2D,
  long  = device to device
}
\DeclareAcronym{DAS}{
  short = DAS,
  long  = distributed antenna system
}
\DeclareAcronym{D-DAS}{
  short = D-DAS,
  long  = dynamic distributed antenna system
}
\DeclareAcronym{DENM}{
  short = DENM,
  long  = decentralized environmental notification message
}
\DeclareAcronym{DSRC}{
  short = DSRC,
  long  = dedicated short range communication
}
\DeclareAcronym{eMBMS}{
  short = eMBMS,
  long  = enhanced multimedia broadcast/multicast service
}
\DeclareAcronym{ETSI}{
  short = ETSI,
  long  = European Telecommunications Standard Institute
}
\DeclareAcronym{FDD}{
  short = FDD,
  long  = frequency division duplex
}
\DeclareAcronym{FD-MIMO}{
  short = FD-MIMO,
  long  = full-dimension MIMO
}
\DeclareAcronym{HetNet}{
  short = HetNet,
  long  = heterogeneous network
}
\DeclareAcronym{ICI}{
  short = ICI,
  long  = inter-carrier interference
}
\DeclareAcronym{IoT}{
  short = IoT,
  long  = Internet of Things
}
\DeclareAcronym{ISI}{
  short = ISI,
  long  = inter-symbol interference
}
\DeclareAcronym{ITS}{
  short = ITS,
  long  = intelligent transport system
}
\DeclareAcronym{LBM}{
  short = LBM,
  long  = leakage based multicast
}
\DeclareAcronym{LTE}{
  short = LTE,
  long  = Long Term Evolution
}
\DeclareAcronym{LTE-V}{
  short = LTE-V,
  long  = LTE for vehicular
}
\DeclareAcronym{MIMO}{
  short = MIMO,
  long  = multiple-input multiple-output
}
\DeclareAcronym{MISO}{
  short = MISO,
  long  = multiple-input single-output
}
\DeclareAcronym{mmW}{
  short = mmW,
  long  = millimetre wave
}
\DeclareAcronym{OFDM}{
  short = OFDM,
  long  = orthogonal frequency division multiplexing
}
\DeclareAcronym{PAPR}{
  short = PAPR,
  long  = peak to average power ratio
}
\DeclareAcronym{PHY}{
  short = PHY,
  long  = physical layer
}
\DeclareAcronym{ProSe}{
  short = ProSe,
  long  = proximity service
}
\DeclareAcronym{PWS}{
  short = PWS,
  long  = public warning system
}
\DeclareAcronym{QoS}{
  short = QoS,
  long  = quality of service
}
\DeclareAcronym{RF}{
  short = RF,
  long  = radio frequency
}
\DeclareAcronym{RoF}{
  short = RoF,
  long  = radio over fibre
}
\DeclareAcronym{RLS}{
  short = RLS,
  long  = recursive least squares
}
\DeclareAcronym{RRH}{
  short = RRH,
  long  = remote radio head\color[rgb]{0,0,0}
}
\DeclareAcronym{beRRH}{
  short = (e)RRH,
  long  = (enhanced) remote radio head\color[rgb]{0,0,0}
}
\DeclareAcronym{eRRH}{
  short = eRRH,
  long  = enhanced remote radio head\color[rgb]{0,0,0}
}
\DeclareAcronym{SAE}{
  short = SAE,
  long  = Society of Automotive Engineers
}
\DeclareAcronym{SC-FDM}{
  short = SC-FDM,
  long  = single-carrier frequency division multiplexing
}
\DeclareAcronym{SINR}{
  short = SINR,
  long  = signal to interference and noise ratio
}
\DeclareAcronym{SNR}{
  short = SNR,
  long  = signal to noise ratio
}
\DeclareAcronym{SON}{
  short = SON,
  long  = self organizing network
}
\DeclareAcronym{SWIPT}{
  short = SWIPT,
  long  = simultaneous wireless information and power transfer
}
\DeclareAcronym{TDD}{
  short = TDD,
  long  = time division duplex
}
\DeclareAcronym{TTI}{
  short = TTI,
  long  = transmission time interval
}
\DeclareAcronym{TXRU}{
  short = TXRU,
  long  = transceiver unit
}
\DeclareAcronym{UMTS}{
  short = UMTS,
  long  = Universal Mobile Telecommunications System
}
\DeclareAcronym{V2I}{
  short = V2I,
  long  = vehicle to infrastructure
}
\DeclareAcronym{V2P}{
  short = V2P,
  long  = vehicle to pedestrian
}
\DeclareAcronym{V2V}{
  short = V2V,
  long  = vehicle to vehicle
}
\DeclareAcronym{V2X}{
  short = V2X,
  long  = vehicle to 'X'
}
\DeclareAcronym{VANET}{
  short = VANET,
  long  = vehicular ad-hoc network
}

\crefformat{chapter}{#2Chapter~#1#3}
\crefformat{lem}{#2Lemma~#1#3}
\crefformat{section}{#2Section~#1#3}
\crefformat{subsection}{#2Section~#1#3}
\crefformat{appendix}{#2Appendix~#1#3}
\crefformat{figure}{#2Figure~#1#3}
\crefformat{table}{#2Table~#1#3}
\crefformat{equation}{#2Equation~(#1)#3}
\crefmultiformat{chapter}{#2Chapters~#1#3}{ and~#2#1#3}{, ~#2#1#3}{, and~#2#1#3}
\crefmultiformat{section}{#2Sections~#1#3}{ and~#2#1#3}{, ~#2#1#3}{, and~#2#1#3}
\crefmultiformat{equation}{#2Equations~(#1)#3}{ and~#2(#1)#3}{, ~#2(#1)#3}{, and~#2(#1)#3}
\crefmultiformat{figure}{#2Figures~#1#3}{ and~#2#1#3}{, ~#2#1#3}{, and~#2#1#3}
\crefmultiformat{table}{#2Tables~#1#3}{ and~#2#1#3}{, ~#2#1#3}{, and~#2#1#3}
\crefrangeformat{equation}{#3Equations~(#1)#4--#5(#2)#6}
\crefrangeformat{figure}{#3Figures~#1#4--#5#2#6}
\crefrangeformat{chapter}{#3Chapters~#1#4 to #5#2#6}

\begin{document}
\title{Cellular Network Architectures for the Society in Motion}
\author{Stefan Schwarz$^\dagger$~\IEEEmembership{Member,~IEEE} and Markus Rupp$^\ddagger$~\IEEEmembership{Fellow,~IEEE}\\ 
$^\dagger$ Christian Doppler Laboratory for Dependable Wireless Connectivity for the Society in Motion\\
$^\dagger\,^\ddagger$ Institute of Telecommunications, Technische Universität (TU) Wien, Austria\\
Email: \{sschwarz,mrupp\}@nt.tuwien.ac.at
\thanks{The financial support by the Austrian Federal Ministry of Science, Research and Economy, by the National Foundation for Research, Technology and Development and by TU Wien is gratefully acknowledged.}
}


\maketitle

\begin{abstract} 
Due to rising mobility worldwide, a growing number of people utilizes cellular network services while on the move. Persistent urbanization trends raise the number of daily commuters, leading to a situation where telecommunication requirements are mainly dictated by two categories of users: 1) Static users inside buildings, demanding instantaneous and virtually bandwidth-unlimited access to the Internet and Cloud services; 2) moving users outside, expecting ubiquitous and seamless mobility even at high velocity. While most work on future mobile communications is motivated by the first category of users, we outline in this article a layered cellular network architecture that has the potential to efficiently support both user groups simultaneously. We deduce novel transceiver architectures and derive research questions that need to be tackled to effectively maintain wireless connectivity for the envisioned Society in Motion.
\end{abstract}

\acuse{beRRH}

\section{Introduction}
\label{sec:Introduction}

Two important global trends are currently challenging many metropolises worldwide: 1) Growing urbanization leads to increased mobility of people commuting to and within cities, regularly overloading public and private transportation systems~\cite{UN_urbanization_2014}; 2) ever-increasing data traffic demands, caused by popular online applications and services, drive cellular networks to their limits~\cite{Ericsson2015}. Since people increasingly utilize their mobile devices for online activities, such as, shopping, entertainment and socializing, while commuting and traveling, these two trends will come together to cause a future challenging situation for wireless communications, where high data rate connectivity must be provided to a large number of potentially highly-mobile users in networks that are already crowded with quasi-static (nomadic) users. Machine-type communication will further aggravate this problem, as road and rail vehicles are expected to employ \ac{5G} mobile networks to provide broadband services to their customers, to support \ac{IoT} applications such as remote sensing and maintenance, and to exchange \ac{ITS} messages to improve road traffic safety and efficiency.

Even though a basic support of few users with speeds as high as 500\,km/h is foreseen in the \ac{4G} \ac{LTE}, the network is not designed to efficiently serve large numbers of high-mobility users. Following the ongoing progress of 5G, it is observed that most research work is motivated by achieving highest peak data rates and large network capacities for quasi-static (nomadic) users, by reducing network latency to enable novel ``tactile Internet'' applications, and by enhancing energy efficiency to decrease the global energy footprint of mobile communications~\cite{5GPPP_2015}. Albeit enhancing mobility is commonly considered a (possibly secondary) 5G goal, it is actually not so much the \ac{PHY} high-mobility support that needs to be improved for the envisioned Society in Motion, but rather the network capacity for very large numbers of mobile users, which is to a large extent a network-level issue. Considering ultra-dense networks that employ small cells “on every lamp-post” to cover areas of mere tens to hundreds of square meters and carrier-frequencies that constantly increase to alleviate bandwidth scarcity, even users with relatively low velocities (say 30\,km/h) have to be considered as highly mobile from a network perspective, further exacerbating the addressed problematic.

\textit{Contribution:}  In this article, we outline and describe a feasible layered cellular network architecture that encompasses many established enabling 5G technologies to not only support commonly agreed 5G targets, but also to enhance network capacity for highly-mobile users. We explain the functions that the individual layers have to implement and how these layers have to cooperate to enable efficient network operation. We furthermore deduce transceiver architectures that play an important key role to realize the envisioned network structure.

\section{Network Architecture}
\label{sec:Net}

\begin{figure*}
  \centering
  \includegraphics[scale = 0.535]{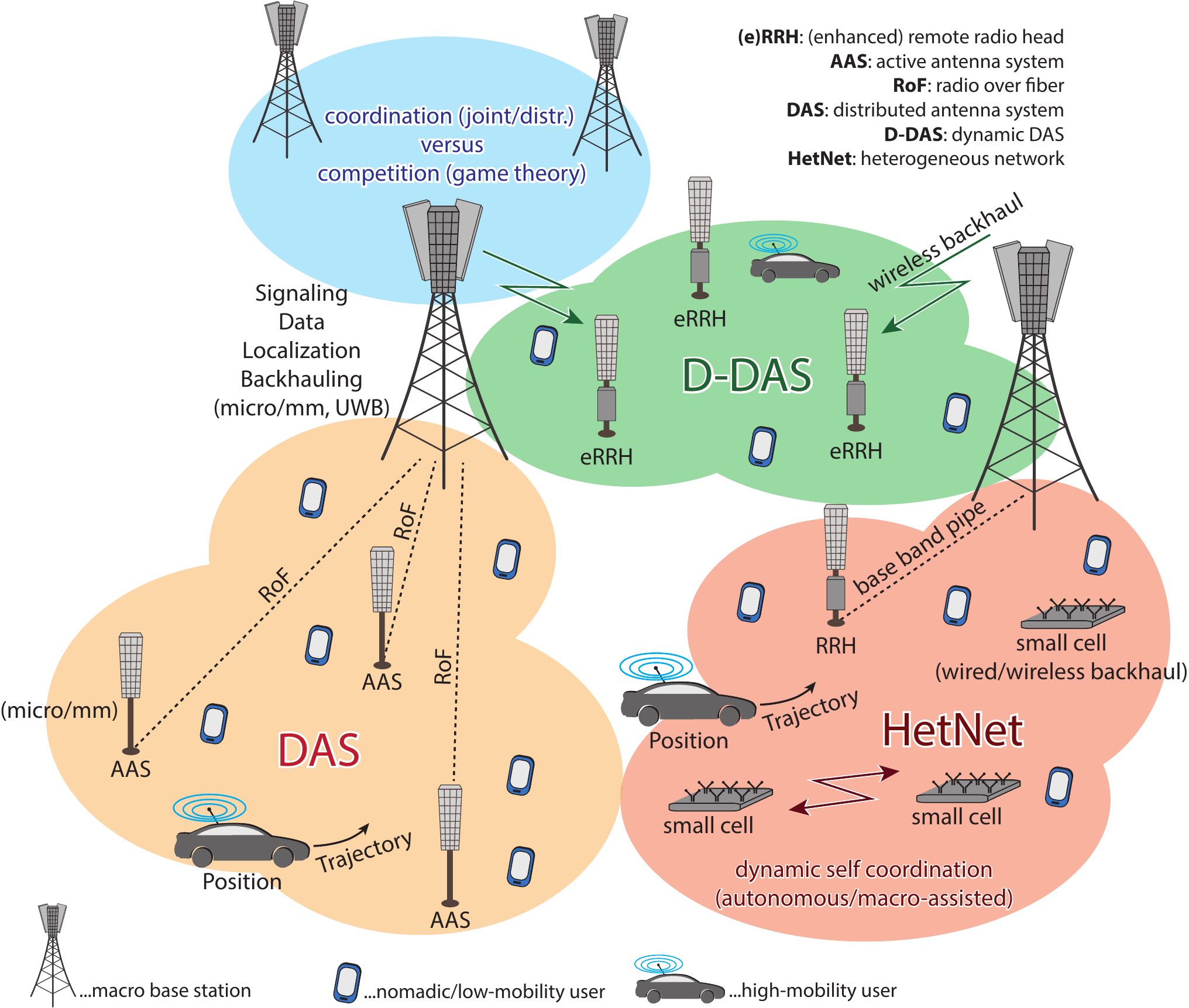}
  \caption{Illustration of a future layered mobile network featuring several types of radio access equipment, to enable efficient support of a large number of highly mobile users in an area crowded with quasi-static (nomadic) users.}
  \label{fig:Net}
\end{figure*}

Macro base stations, when operating in the lower micro wave frequency bands (e.g., 800\,MHz - 2\,GHz), are the method of choice for providing coverage over large geographical areas with comparatively small number of base station sites. Hence, such macro base stations will also form the backbone of future \ac{5G} mobile networks as illustrated in~\cref{fig:Net}. It has, however, already been recognized during the development of \ac{4G} mobile networks, that macro base stations alone can neither maintain increasing network capacity demands, nor can they support energy efficiency requirements imposed on \ac{5G} systems~\cite{5GPPP_2015}. 

Thus, network densification was a central theme in the development of \ac{4G} cellular networks and will continue being so in \ac{5G}, in order to sustain ever-increasing network capacity demands~\cite{Damnjanovic2011}. In \ac{4G} \acp{HetNet}, mostly autonomous small cells are employed to provide coverage and capacity in indoor and hot-spot locations. Small cells, however, are not a satisfactory solution for supporting mobile users, since coverage areas are small, implying frequent time-consuming and error-prone hand-overs between different autonomous cells. This has been recognized by the \ac{3GPP} and countermeasures have been taken by including dual-connectivity within \ac{LTE} Release\,12. Yet, this a-posteriori fix is not the most efficient and reliable solution, because the intrinsic macro-diversity provided by multiple neighboring small cells cannot be exploited, since small cells do not support advanced joint transmission techniques due to the lack of a powerful backhaul. 

Hence, \acp{DAS} will in part outstrip small cells in \ac{5G} especially for professional indoor micro \ac{C-RAN} solutions~\cite{Checko2015} and for outdoor network capacity enhancement, as they facilitate advanced \ac{CoMP} schemes that improve network capacity~\cite{Schwarz-TWC2014}. The major disadvantage of a conventional \ac{DAS} is that its component \acp{AAS} require dedicated \ac{RoF} (or equivalent) infrastructure to communicate with their controlling macro base station. 

To alleviate this drawback of \acp{DAS}, we propose a novel type of network access node, the \ac{eRRH}, which is a \ac{RRH} featuring an \ac{AAS} and a wireless backhaul to the macro base station as detailed in~\cref{sec:eRRH}. Such \acp{eRRH} can be placed opportunistically within the network, requiring power supply only. Macro base stations can take control over \acp{eRRH} using wireless backhaul connections. This allows establishing \acp{DAS} on-demand by associating a number of \acp{eRRH} with one or several macro base stations; such a \ac{D-DAS} is illustrated in the upper-right part of~\cref{fig:Net}. High capacity wireless backhauling can be achieved with doubled-sided massive \ac{MIMO} systems, enabling highly directive transmission between macro base station and \acp{eRRH} as further described in~\cref{sec:double_massive}. Macro base stations coordinately associate \acp{eRRH} amongst each other, to achieve certain capacity, diversity and/or mobility requirements of their users.   

To efficiently support \ac{5G} requirements, the individual layers of the presented heterogeneous cellular network architecture have to collaboratively fulfill certain tasks as we describe below in detail. In Table~\ref{fig:Coord} we summarize the three levels of cooperation required in the described layered cellular network architecture to enable realization of \ac{5G} targets. Cooperation on the macro-layer will involve enhanced \ac{SON} features that apply Game theoretic and other optimization methods to achieve large scale network coordination over long time scales (minutes to hours). Cooperation in-between the macro- and micro-layers is performed on the meso-level, which mostly deals with user and network node assignment issues that apply over areas covered by several macro base stations. Micro-level cooperation, finally, utilizes advanced multi-antenna and multi-point transmission/reception techniques on a \ac{TTI} time scale, to achieve efficient and robust coordination of few neighboring radio access nodes of the micro-layer.        

\begin{figure*}
	\renewcommand\figurename{Tab.}
	\setcounter{figure}{0} 
	\renewcommand{\thefigure}{\Roman{figure}}
  \centering
  \includegraphics[scale = 0.535]{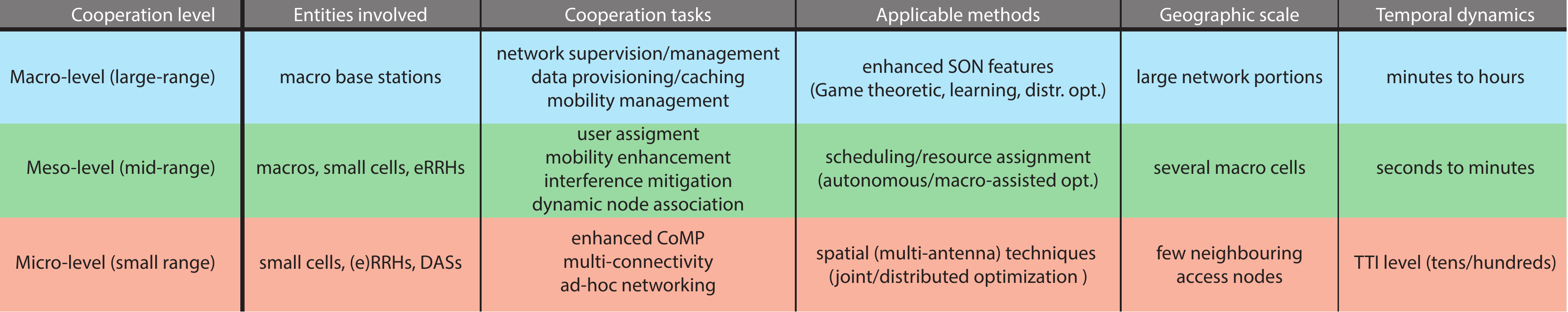}
  \caption{Table describing levels of cooperation required on and in-between different layers of the network illustrated in~\cref{fig:Net}.}
  \label{fig:Coord}
\end{figure*}

\subsection{Macro-Layer Functions}
\label{sec:macro}

In future cellular networks, the importance of macro base stations for serving users will decrease. Nevertheless, the macro-layer will play a central role in \ac{5G} mobile communications, by undertaking network management functionality and providing wireless backhaul to network access nodes of the micro-layer, such as, small cells and \acp{RRH}. We anticipate that future macro base stations will be employed to supervise and maintain data transmissions from the micro-layer, thereby enhancing transmission efficiency by utilizing the holistic network view available at the macro-layer. For that purpose we identify the following key functions of the macro-layer: 

\subsubsection{Mobility management and user assignment} A major task of the macro-layer is in assisting the assignment of users amongst network access nodes of the micro-layer in the long term, respectively over large geographic areas. This is especially important for highly-mobile users that traverse quickly through coverage regions of small cells. Depending on the sojourn time of coverage regions, it can be beneficial to avoid the attachment of highly-mobile users to small cells, in order to minimize signaling load. If macro base stations are able to determine locations and trajectories of users (e.g., through explicit user feedback or employing multi-antenna localization techniques), they can also assist in reserving resources at network access nodes along the path of users to minimize hand-over interruptions. These methods are especially promising for users that move along well-defined paths such as (rail-)roads.   
   
\subsubsection{Data provisioning and geocasting} Data provisioning can help reducing backbone load and latency, by caching popular content, such as videos, as close to users as possible~\cite{Bastug2015}. Here, macro base stations can support identification of user hot-spot locations to initiate data caching at the respective small cells of the micro-layer. In the context of mobile users, data provisioning goes hand-in-hand with mobility management as described above to minimize latency after hand-overs. Furthermore, in applications such as \ac{ITS}, information is mostly relevant only for users in certain geographic areas; e.g., road hazard warnings are important for users moving on the respective stretch of road. Such schemes can efficiently be implemented if geocasting is supported by the macro-layer.

\subsubsection{Provisioning of signaling information} Even though payload data transmission will mostly be handled by the micro-layer in future mobile networks, transmission of signaling information may still reside with the macro-layer. Such control-plane/user-plane splitting concepts promise robust and efficient wireless connectivity for mobile users in \acp{HetNet}~\cite{Ishii12}. Especially with \ac{mmW}-based small cells, which are prone to signal outage~\cite{Rangan2014}, keeping the control-plane at the macro-layer can reduce loss of connection by enabling fast hand-over to alternative access nodes.

\subsubsection{Coverage backup} The macro-layer will retain its role as reliable coverage solution in areas that do not economically justify provisioning of high performance micro-layers (e.g., sparsely populated rural areas and back country). Furthermore, the macro-layer will provide back-up connectivity whenever network access nodes of the micro-layer are in outage, thus enhancing reliability of data transmission through extra macro diversity. This is especially important when employing \ac{mmW} based radio access on the micro-layer, since outage probability in the \ac{mmW} regime is high due to severe shadowing effects~\cite{Rangan2014}. Moreover, the macro-layer will support users at highest mobility that traverse coverage regions of the micro-layer in very short time (fractions of a second) and cannot efficiently be co-scheduled with other users on the micro-layer for the following reason: The high-performance micro-layer will apply aggressive spatial multi-user multiplexing to efficiently serve large number of users in parallel. Such schemes require accurate \ac{CSIT}, which is commonly not available at very high mobility due to fast temporal channel variations. Hence, micro-layer efficiency can be enhanced by offloading highest mobility users to the macro network. Further efficiency improvements for highest mobility users that move along predetermined paths (e.g., rail-roads) are possible by providing macro base stations with dedicated distributed antennas to reduce the access distance  between users and antenna arrays~\cite{Mueller2015}.

\subsection{Meso-Layer Functions}
\label{sec:meso}

The meso-layer covers the interaction between macro base stations and other (semi)-autonomous radio access nodes of the micro-layer, such as, small cells and \acp{eRRH}. For that purpose it has to support the following functions: 

\subsubsection{Wake-up on-demand} An important \ac{5G} goal is the reduction of the global energy footprint of mobile networks. This can most easily be achieved by deactivating network nodes of the micro-layer whenever they are not required to satisfy demand, as gauged by the macro-layer. Largest energy savings are possible when \ac{RF} chains of deactivated access nodes can be completely powered down; yet, this implies power up delay when reactivating, which can be problematic for highly-mobile users. To avoid excessive delay, the macro-layer should hence apply predictive methods to determine early on when to reactivate radio access nodes.       

\subsubsection{Dynamic access node association} The meso-layer is responsible for associating \acp{eRRH} to macro base stations to form \acp{D-DAS} on-demand. This implies coordination amongst macro base stations to determine optimal \ac{eRRH} associations, as well as forwarding the corresponding control information to the micro-layer. With \acp{D-DAS} the trailing cell concept can be effectively realized, which virtually moves the signal of a cell along with users by switching between \acp{eRRH}.\footnote{Notice, in literature this is also known as moving cell; we avoid this notation,
since it is also used for small cells that are mounted on vehicles.} 

\subsubsection{Management of coordination areas} The meso-layer has to identify small cells of the micro-layer that should employ \ac{CoMP} techniques to improve performance (reduce interference, enhance macro-diversity, improve mobility). This implies assisting the dynamic formation of network coordination areas and forwarding of coordination information in-between small cells as well as between the macro- and micro-layers, in case of macro-assisted coordination and control-plane/user-plane splitting.  

\subsubsection{Wireless backhauling} Future \ac{5G} networks will employ highly directive beamforming, as enabled by massive \ac{MIMO} and \ac{mmW} technologies, to establish on-demand wireless backhaul connections between network access nodes of the same layer and across layers (macro to micro). Since these network access nodes are basically static, accurate \ac{CSI} can be obtained at all involved nodes with minimal effort, facilitating sophisticated coordinated beamforming/precoding techniques. On-demand wireless backhauling extends the available options for opportunistic placement of network access nodes, since fixed infrastructure requirements are minimized. It also enables advanced dynamic coordination of network access nodes that are not equipped with powerful fixed backhaul.

\subsection{Micro-Layer Functions}
\label{sec:micro}

The micro-layer is the actual user access layer of the mobile network. It has to support network capacity and other requirements imposed on \ac{5G} systems. This will necessitate on-demand and dynamic \ac{TTI} level coordination of multiple neighboring radio access nodes to control interference, provide robustness with respect to signal outages and enhance mobility support. The corresponding cooperation schemes will involve comparatively simple time-frequency coordination of resource allocations but also sophisticated spatial \ac{CoMP} techniques, such as, joint transmission from spatially distributed access nodes and interference alignment. Coordination amongst micro-layer access nodes can either be autonomous, employing distributed optimization algorithms such as reinforcement learning~\cite{Simsek15}, or macro-assisted, enabling joint optimization at a central entity. With such methods, the micro-layer can efficiently and dependably realize (amongst others) the following functions for mobile users:

\subsubsection{User association} In current \ac{4G} cellular networks, user association to network access nodes is basically triggered from the user-side based on channel quality measurements. To avoid overloading of crowded small cells and base stations, several strategies for autonomous and coordinated load balancing in \acp{HetNet} have been proposed in literature~\cite{Ye2013}, often applying SNR thresholding to artificially increase the coverage areas of lightly loaded cells. Even though such methods can produce favorable load conditions in static scenarios, with highly-mobile users user association should additionally account for sojourn times of coverage regions to avoid large signaling overhead. Highly-mobile users commonly move along specific paths (roads); this repetitive side information can be utilized by the access nodes to adaptively train optimal radiation beam-patterns and user association strategies over time, e.g., through reinforcement learning techniques.      

\subsubsection{Provisioning of spatial macro-diversity} The micro-layer of future ultra-dense cellular networks will inherently possess a large degree of spatial macro-diversity. By means of dynamic coordination amongst radio access nodes, this macro-diversity can be made available to users to enhance robustness with respect to signal outages. In current mobile networks, however, backhaul connections between small cells are mostly not sufficiently powerful to enable fast autonomous coordination, as required for mobile users. This problem can be alleviated through wireless backhauling as described in~\cref{sec:meso}. 

\subsubsection{Vehicular communications} In the context of high-mobility users, \acp{VANET} have gained interest in recent years for \ac{ITS} applications, promising enhanced road traffic safety and efficiency. Lately the interest in mobile communications for such use cases is increasing, since suitable technology is comparatively cheaply available off-the-shelf and the required network infrastructure is practically ubiquitous. This has also been recognized by the \ac{3GPP} and early LTE standardization efforts for vehicular (LTE-V) are ongoing in the development of Release\,14, with the goal of supporting \ac{ITS} services and providing high-bandwidth infotainment applications for in-car users. We anticipate that \ac{5G} networks will utilize a combination of \acp{VANET} and mobile communications to mutually enhance the broadband experience of participating users.


\section{Deduced Transceiver Architectures}
\label{sec:TX}

To realize the network architecture illustrated in~\cref{fig:Net} several new or modified types of radio access networks have to be treated, whose specifics and features we discuss below.

\subsection{Double-Sided Massive MIMO}
\label{sec:double_massive}

\begin{figure}
  \centering
  \includegraphics[scale = 0.45]{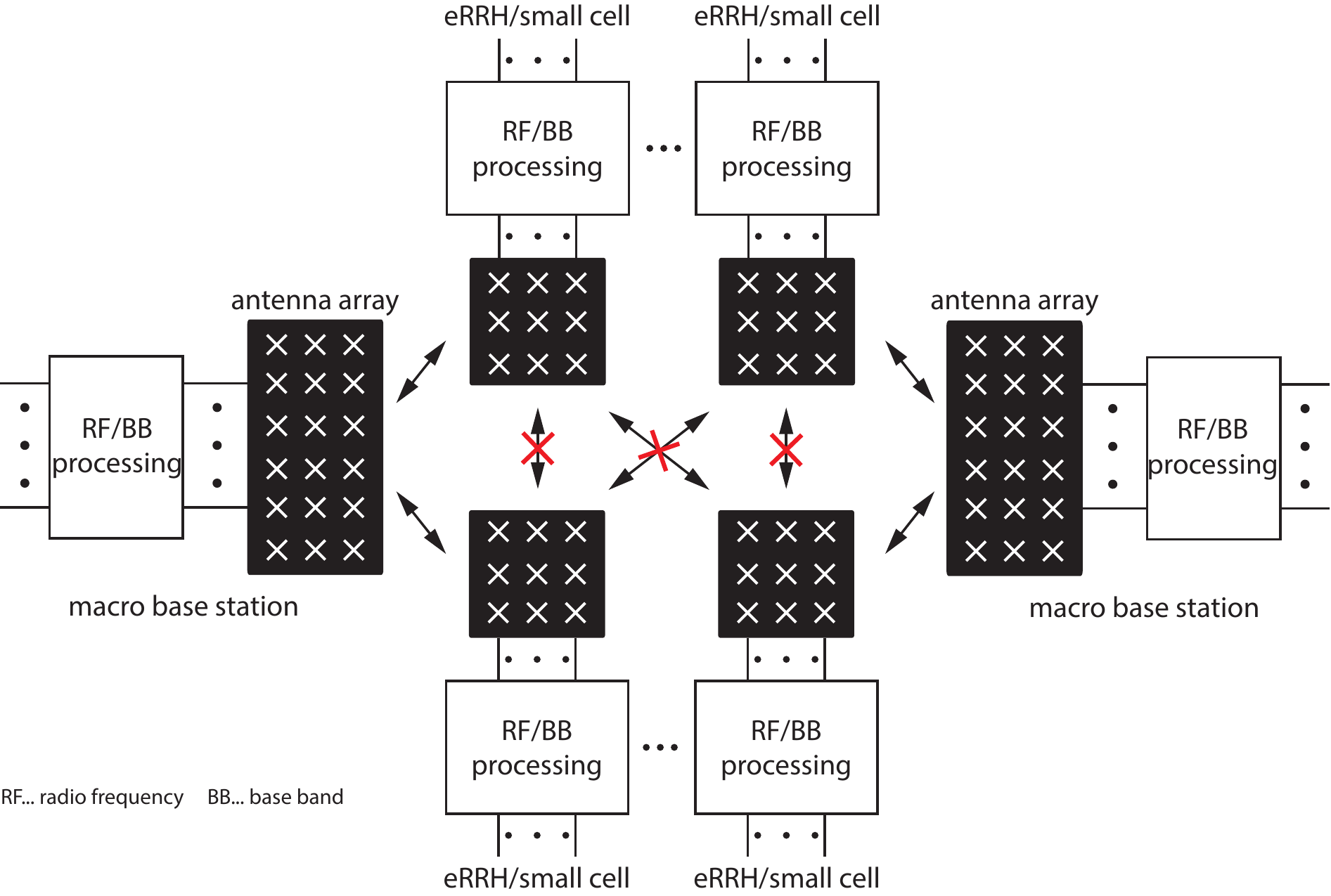}
  \caption{Double-sided massive MIMO system describing the wireless backhaul link between macro base stations and small cells/eRRHs.}
  \label{fig:Double}
\end{figure}

Wireless backhauling between macro base stations and small cells/\acp{eRRH}, as described in~\cref{sec:meso}, has to provide very high capacity, in order to sustain high rate data transmissions to several users of the micro-layer in parallel. Therefore the \ac{mmW} band lends itself for these wireless connections, due to the large amount of available bandwidth in the \ac{mmW} regime. Alternatively, since \ac{mmW} transmission is sensitive to weather conditions (rain, fog) and thus prone to outage, transmission in the low GHz regime can be applied to reliably cover large distances. In both cases, large antenna arrays on both ends of the links, i.e., at macro base stations and small cells/\acp{RRH}, make sense to achieve large spatial multiplexing gains and guarantee minimal interference to unintended network nodes. This leads to so-called double-sided massive \ac{MIMO} transmission, as illustrated in~\cref{fig:Double}. Such systems are to date to a large extend unexplored; only in~\cite{Schwarz_CCNC16} the authors investigate several downlink multi-user \ac{MIMO} transceivers for double-sided massive antenna arrays, showing that the achieved rate is very sensitive to antenna correlation, especially at the transmitter-side. Double-sided massive \ac{MIMO} systems are also relevant in \ac{SWIPT}, due to the large beamforming gains achievable at transmitter and receiver that enable comparatively efficient remote energy supply.  

In~\cref{fig:Double}, we consider an example of wireless backhauling between macro base stations and several small cells/\acp{eRRH} over a dedicated bandwidth. Since all involved network nodes are static, the wireless channels in-between them will not vary too much over time, allowing to obtain accurate estimates of \ac{CSI} at the transmitters and receivers with minimal effort. The corresponding system model is equivalent to interfering multi-cell data transmissions in cellular networks, despite having massive antenna arrays on both ends of the links. Yet, transceiver designs cannot directly be adopted from such cellular systems, since, for reasons of complexity, in massive \ac{MIMO} hybrid base band/\ac{RF} signal processing is applied in many cases to reduce the number of required \ac{RF} chains~\cite{Alkhateeb2014}. In case wireless backhauling employs the same bandwidth as data transmission between users and access nodes of the micro-layer, the system model of~\cref{fig:Double} has to be further extended to enable joint optimization of backhauling and user data transmission. This corresponds to a situation with mixed single-/double-sided massive \ac{MIMO} transmission, since user equipment will in general not support massive antenna arrays for reasons of space and complexity. Hence, such systems open a new avenue of research and engineering tasks, ranging from practical transceiver designs to fundamental questions regarding limits of achievable data rates of double-sided massive \ac{MIMO}.     

\begin{figure}
  \centering
  \includegraphics[scale = 0.45]{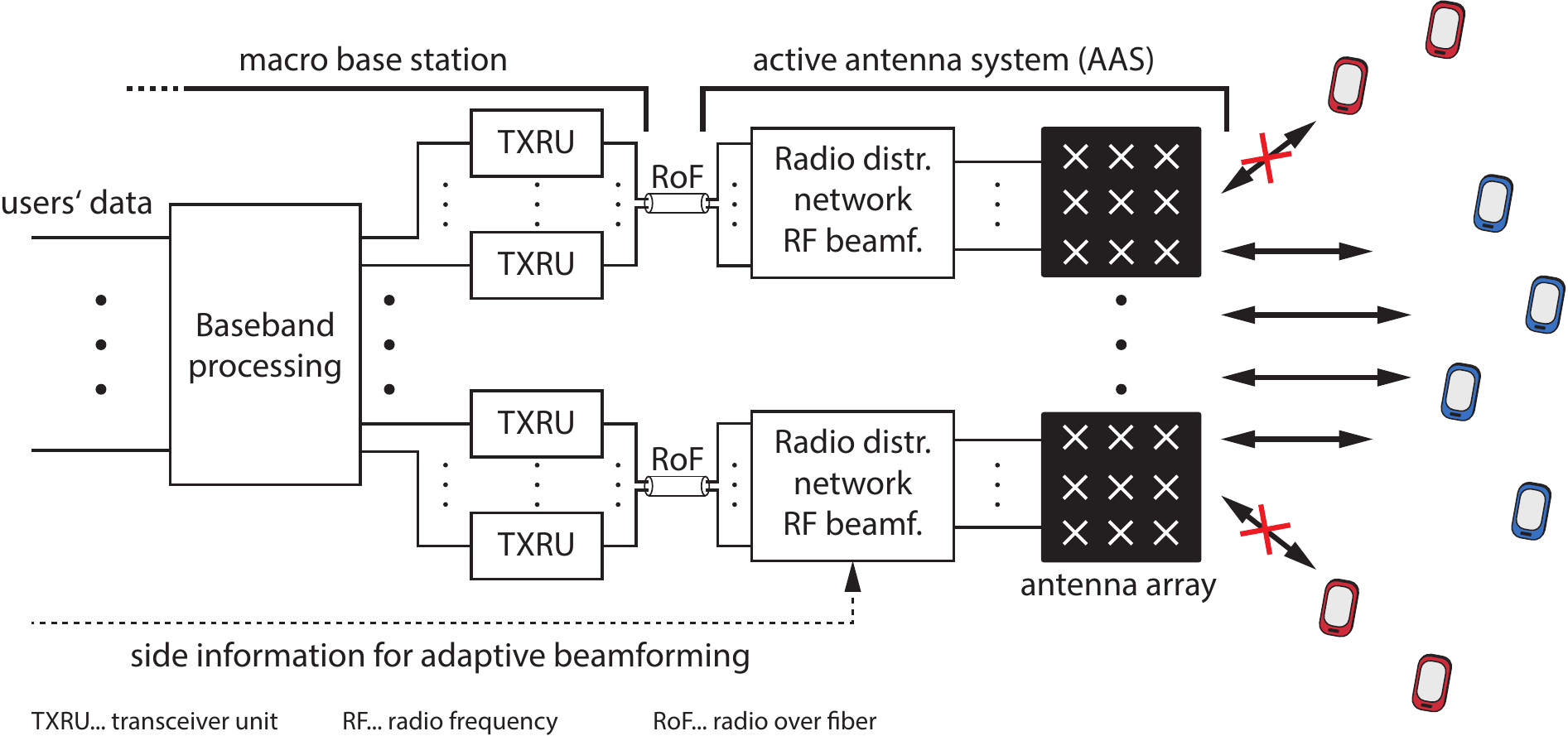}
  \caption{Massive distributed antenna systems, featuring multiple active antenna arrays that are distributed over a certain geographical region and are connected to a macro base station employing radio of fibre transmission.}
  \label{fig:DAS}
\end{figure}

\subsection{Massive Distributed Antennas Arrays}

DASs are well known to improve coverage of cellular networks, to reduce signal outages through additional macro-diversity and to enhance network capacity~\cite{Heath2011}. In contrast to autonomous small cells, \acp{DAS} have the advantage that they are centrally controlled by a macro base station, which enables the implementation of highly efficient joint transmission and coordinated beamforming techniques~\cite{Schwarz-TWC2014}. On the other hand, if such distributed antennas are well isolated of each other, they can also be utilized to radiate independent signals mimicking the behaviour of autonomous small cells. An efficient way of implementing a \ac{DAS} is to couple several spatially distributed \acp{AAS} to a macro base station utilizing high-bandwidth and low-latency \ac{RoF} technology. Employing large antenna arrays at the \acp{AAS} additionally allows harvesting the gains promised by three-dimensional beamforming and \ac{FD-MIMO} solutions, techniques that are currently of large interest within the scientific community as well as standardization of \ac{LTE} Release\,14 and beyond. 

A simplified block-diagram of such massive \ac{DAS} is illustrated in~\cref{fig:DAS}; here, the macro base station is responsible for processing the users' signals and for up- and down-converting them to and from \ac{RF}. The \acp{AAS} enable adaptive \ac{RF} beamforming through radio distribution networks, consisting of phase shifters and \ac{RF} combiners, that are remotely controlled by the macro base station over dedicated side links. Massive \acp{DAS} can support highest network capacity, by coordinating \ac{FD-MIMO} transmissions from several massive antenna arrays to achieve large spatial multiplexing gains with minimal interference. Compared to autonomous small cell solutions, they additionally enhance the dependability of the wireless connection, since users can, in case of outage, be immediately and seamlessly switched over from one \acp{AAS} to the next, without requiring any coordination amongst independent network access nodes. This feature is also advantageous for achieving reliable and robust support of high-mobility users~\cite{Mueller2015}, since frequent hand-overs between autonomous access nodes can be avoided. 

\subsection{Enhanced Remote Radio Heads}
\label{sec:eRRH}

The classical \ac{DAS}, as already utilized nowadays in cellular networks, has three main disadvantages: 1) the assignment of \acp{AAS} to base stations is static;  2) dedicated high-bandwidth low-latency links, such as \ac{RoF}, are required to attach the active antennas to the base station; 3) all signal processing has to be performed by the base station, since \acp{AAS} do not possess any processing capabilities. The first two drawbacks can be avoided by extending the \acp{AAS} with \ac{RF} chains to act as \acp{RRH} and by attaching these \acp{RRH} on-demand dynamically to macro base stations to form \acp{D-DAS} as required. To alleviate the third drawback, we consider enhancing the capabilities of \acp{RRH} to enable autonomous base band and/or \ac{RF} beamforming. Equipped with such \acp{eRRH}, signal processing load at the base stations is reduced to multi-user scheduling and resource assignment, whereas base band precoding and \ac{RF} beamforming is automatically optimized by the \acp{eRRH}. 

\begin{figure}
  \centering
  \includegraphics[scale = 0.45]{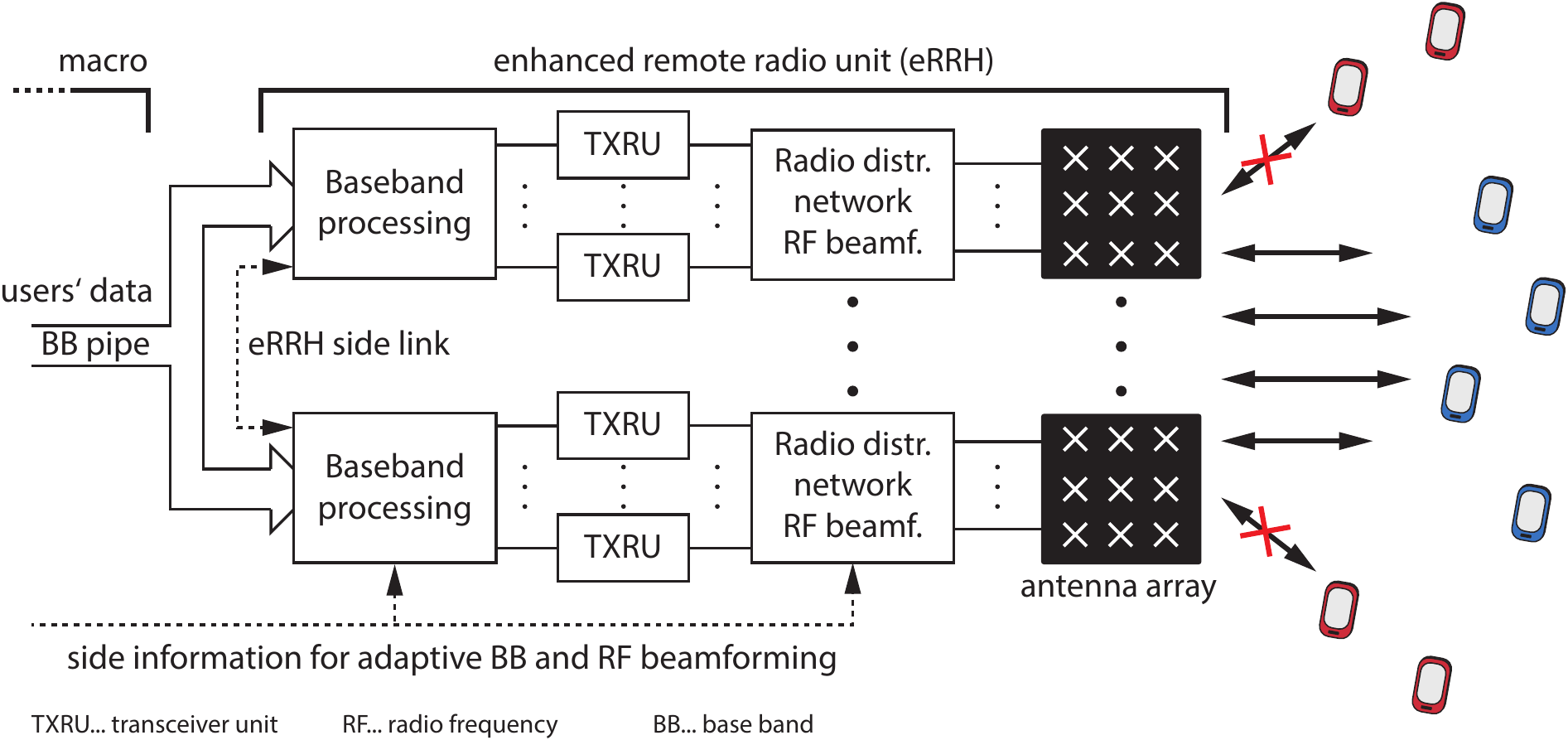}
  \caption{Dynamic distributed antenna system, utilizing enhanced remote radio heads to enable macro-assisted and coordinated base band and radio frequency beamforming at several spatially distributed network access nodes.}
  \label{fig:RRH}
\end{figure}

In~\cref{fig:RRH}, we show the block-diagram of such a \ac{DAS} employing \acp{eRRH}: the pre-processed base band data of the users is forward to the \acp{eRRH} either over dedicated connections or sharing the bandwidth with other transmissions. To optimize the beamforming/precoding weights, the \acp{eRRH} are provided with side-information about intended and unintended users by the base station. Although similar to classical decode-and-forward relay nodes, \acp{eRRH} thus obtain additional input information from the macro base station and can coordinate amongst each other to enable joint optimization of beamforming/precoding weights. Compared to traditional \acp{RRH}, precoding with \acp{eRRH} occurs entirely transparent to base stations, allowing to reduce the required backhaul capacity, since the actual user data is in general of much smaller dimension than the precoded signal especially with massive \ac{MIMO}. This is highly attractive for \ac{D-DAS} systems employing massive antenna arrays, since such system share the wireless medium as backhaul connection. Furthermore, keeping beamforming/precoding adaptation as close to the wireless channel as possible enables reducing the latency, because the up- and downlink delay of \ac{CSI}-sharing between base station and \acp{eRRH} is omitted. This makes such solutions suitable for high-mobility scenarios that require fast beamforming adaptation due to strong temporal channel variations.

\section{Research Challenges}
\label{sec:chall}

Utilization of the presented layered cellular network architecture to support the envisioned Society in Motion comes with several research challenges, associated with the interaction of network access nodes within and in-between layers. We summarize the most critical research issues below:

\paragraph{Network coordination} Dynamic coordination of network access nodes on all layers of the network is a central enabler for the features described in \cref{sec:Net}. A static or semi-static setup of network parameters, possibly even by hand as still common practice in current cellular networks, is neither feasible nor efficient in future ultra-dense \ac{5G} mobile communications; networks should rather support autonomous self-optimization. For that purpose, efficient and reliable coordination methods for the different network layers need to be developed. On a small scale, involving several network access nodes, cooperative schemes that jointly optimize the state of involved nodes enable highest performance; yet, such schemes do not scale well with the number of involved nodes. Thus, on a large scale, competitive (Game theoretic) schemes are promising feasible candidates for achieving efficient coordination. Especially the interaction of such cooperative and competitive schemes needs to be further addressed by research. 

\paragraph{Wireless backhauling solutions} On-demand wireless backhauling is in many cases an enabler for sophisticated coordination schemes, since a powerful wired backhaul can often not be supplied without significant cost, specifically in outdoor locations. However, this requires very high capacity solutions for wireless backhauling to support the data rate demands of many attached users. Such solutions may be available in \ac{mmW} and massive \ac{MIMO} technologies, utilizing large antenna arrays on both ends of the transmission link. Transceiver architectures and computationally efficient \ac{CoMP} schemes for such double-sided massive \ac{MIMO} systems, enabling efficient and reliable backhauling, need to be developed.   

\paragraph{\ac{PHY} enhancements} When serving high-mobility users, the \ac{PHY} itself is often the limiting factor of achievable rate. Several studies have identified severe weaknesses of \ac{LTE} when serving users at high velocity. These are mostly caused by inaccurate \ac{CSI} at transmitter and receiver, hindering the application of sophisticated transmission and reception schemes. Enhancements of pilot symbol structures and \ac{CSI} feedback algorithms have been proposed, which trade-off increased overhead for reduced \ac{CSI} inaccuracy. To get the most out of such schemes, they should operate adaptively with respect to the delay-Doppler dispersion characteristics of the wireless channel. With \ac{5G}, novel \ac{PHY} multi-carrier waveforms are likely to be employed; this gives the opportunity to design waveform parameters to optimally match the dispersion characteristics of the channel. To efficiently and robustly support indoor/outdoor static/mobile users with diverse channel characteristics, \ac{5G} waveforms should be able to adjust time-frequency spacing, the applied prototype pulse as well as the \ac{TTI} length, potentially for each user individually.

\section{Conclusion}
In this article, we discussed challenges for mobile communications imposed by a likely future development, where a large number of highly-mobile users must efficiently and dependably be served in parallel to even more quasi-static users. We outlined a layered cellular network architecture and discussed functions that must be realized by the individual layers of the network to support such demands. We highlighted the importance of dynamic and autonomous self-coordination to achieve efficient operation of the network and we identified several research challenges that need to be addressed to realize a future connected Society in Motion.

\bibliographystyle{IEEEtran_no_url}

\begin{footnotesize}
\bibliography{Network_SiM}
\end{footnotesize}

\end{document}